\documentclass[journal]{IEEEtran}

\usepackage{graphicx}
\usepackage{amsmath}
\usepackage{amsfonts}
\usepackage{algorithm}
\usepackage{algorithmic}
\usepackage{makecell}
\usepackage{booktabs}
\setcellgapes{3pt}
\usepackage{multirow}
\setcellgapes{3pt}

\usepackage[justification=centering]{caption}
\usepackage{color}
\usepackage{verbatim}

\usepackage{cite}

\ifCLASSINFOpdf
\else
\fi
\begin{document}
%
\title{Distributed Resource Scheduling for Large-Scale MEC Systems: A Multi-Agent Ensemble Deep Reinforcement Learning with Imitation Acceleration}

\author{Feibo Jiang, Li Dong, Kezhi Wang, Kun Yang and Cunhua Pan
	\thanks{
		Feibo Jiang (jiangfb@hunnu.edu.cn) is with Hunan Provincial Key Laboratory of Intelligent Computing and Language Information Processing, Hunan Normal University, Changsha, China, Li Dong (Dlj2017@hunnu.edu.cn) is with Key Laboratory of Hunan Province for New Retail Virtual Reality Technology, Hunan University of Commerce, Changsha, China, Kezhi Wang (kezhi.wang@northumbria.ac.uk) is with the department of Computer and Information Sciences, Northumbria University, UK,   Kun Yang (kunyang@essex.ac.uk) is with the School of Computer Sciences and Electrical Engineering, University of Essex, CO4 3SQ, Colchester, UK, Cunhua Pan (Email: c.pan@qmul.ac.uk) is with School of Electronic Engineering and Computer Science, Queen Mary University
		of London, London, E1 4NS, UK.}
}

\markboth{Submitted for Review}%
{Shell \MakeLowercase{\textit{et al.}}: Bare Demo of IEEEtran.cls for IEEE Journals}
%



\maketitle

\begin{abstract}

We consider the optimization of distributed resource scheduling to minimize the sum of task latency and energy consumption for all the Internet of things devices (IoTDs) in a large-scale mobile edge computing (MEC) system. 
To address this problem, we propose a distributed intelligent resource scheduling (DIRS) framework, which includes centralized training relying on the global information and distributed decision making by each agent deployed in each MEC server. 
More specifically, we first introduce a novel multi-agent ensemble-assisted distributed deep reinforcement learning (DRL) architecture, which can simplify the overall 
neural network structure of each agent by partitioning the state space and also improve the performance of a single agent by combining decisions of all the agents. Secondly, we apply action refinement to enhance the exploration ability of the proposed DIRS framework, where the near-optimal state-action pairs are obtained by a novel Lévy flight search. Finally, an imitation acceleration scheme is presented to pre-train all the agents, which can significantly accelerate the learning process of the proposed framework through learning the professional experience from a small amount of demonstration data. Extensive simulations are conducted to demonstrate that the proposed DIRS framework is efficient and outperforms the existing benchmark schemes.



\end{abstract}


%
\IEEEpeerreviewmaketitle

\section{Introduction}
Recently, with the rapid increase of resource-intensive tasks, e.g., augmented reality (AR), Internet of things (IoT) applications and autonomous driving, the quality of our life has the potential to be improved greatly. However, due to the limited size and battery life of IoT devices (IoTDs), these applications may be difficult to be implemented in practice. Fortunately, mobile edge computing (MEC) has been proposed recently as a promising technique to liberate IoTDs from computation-intensive tasks by allowing them to offload their high workloads to edge servers \cite{mao2017survey}.

However, due to the large number of IoTDs, one edge server may not be powerful enough to support all the devices at the same time. Thus, multiple MECs may be deployed to support the IoTDs. Then, it is critical to determine the offloading decision and resource allocation between computing and communication resource from different MECs to IoTDs. \cite{mao2017survey}.

Unfortunately, the above problem is normally formulated as a mixed integer nonlinear programming (MINLP) problem \cite{jiang2019deep}, which involves integer variables (i.e.,offloading decision) and continuous variables (i.e., communication and computing resource allocation). This problem is very difficult to solve in general, especially in large-scale IoTD scenarios and dynamic environments. Some traditional solutions have been applied to solve the above problem, such as  game theory\cite{liu2017decentralized}, branch-and-bound method\cite{narendra1977branch} and dynamic programming\cite{bertsekas1995dynamic}. However, these solutions normally needs a large amount of computing resource and it is difficult to realize online decision making process. Some other solutions, such as convex relaxation\cite{dinh2017offloading} and heuristic local search\cite{bi2018computation} algorithms are also applied to handle the above problems. However, those algorithms normally need a considerable amount of iterations to achieve a satisfying local optimum, which may not be suitable for dynamic environment.  
Moreover, with the increase of the number  of IoTDs and MEC servers, the complexity of the above-mentioned traditional solutions increases significantly, which makes them very difficult to be applicable in large-scale environment.

Fortunately, the emerging deep reinforcement learning (DRL) approach has shown great potential in solving the above-mentioned joint optimization problem. However, there are still several challenges yet to be addressed: 1) Value-based DRL (e.g., Q-earning\cite{liu2017enb}, DQN\cite{min2019learning} \cite{he2017integrated} and double DQN \cite{chen2018optimized}) can only work well in a limited action space, which is inefficient in large-scale application scenarios; 2) Although policy-based DRL (e.g., DDPG\cite{liu2018energy} \cite{jiang2020stacked}, A3C \cite{feng2019cooperative}) can update policy by computing policy gradient for maximizing the expected value of Q function, it is difficult to converge in dynamic environment, especially for distributed architecture \cite{lowe2017multi}.

In this paper, we aim to develop a distributed intelligent resource scheduling (DIRS) framework, which can be applied to large-scale MEC systems with multiple MECs and IoTDs in dynamic environment. To enhance the performance of the framework, 
centralized training scheme is designed, whereas decentralized decision making is proposed to increase the flexibility of the framework. The main contributions are summarized as follows:

Firstly, the system model is proposed with the aim of minimizing the sum of task latency plus energy consumption for all the IoTDs. Then, we decompose the proposed MINLP optimization problem into an offloading decision sub-problem and a communication and computing resource allocation sub-problem, which can reduce the complexity of the original problem and guarantee that the solutions meet all the constraints.

Then, we propose a distributed multi-agent ensemble DRL framework for solving the decision making sub-problem. In this framework, one agent is deployed in each MEC to conduct the distributed decision making for IoTD offloading tasks to this MEC. To improve the  performance of the whole system, we have a scheduler deployed in the Core-MEC (C-MEC) to conduct the centralized training with the global information in the training stage. Ensemble learning is introduced into this distributed DRL framework to simplify the neural network structure of each agent by partitioning the state space and improve the performance of a single agent by combining decisions of all the agents. Once the centralized training is completed, each agent in MEC can make decentralized offloading decision only with local information.
 
Next, we present a Lévy flight search as the action refinement to find the best actions for the DRL model according to the current state. Lévy flight search can help the DRL framework to skip the local optimum. In the Lévy flight search, h mutation operator is used to generate mutant vector according to the channel state information, and then the Lévy crossover operator is applied to avoid candidate vector trapping into the local optimum. Finally, greedy selection operator is applied to select the better solution between  candidate vector and original solution.
 
Finally, we propose an imitation acceleration scheme, which combines the DRL and imitation learning to accelerate the training process of all the agents. We first generate a small amount of demonstration data, and then we pre-train all the agents using a novel demonstration loss function, which can reduce the training time and increase the stability of the DRL based framework.
 
The rest of this paper is organized as follows. Section II presents a review of related works. Section III describes the system model and problem formulation. Section IV introduces two different distributed DRL paradigms. Section V describes the detailed design of the DIRS framework. Section VI presents the simulation results, followed by the conclusions in Section VII.
 
 \section{Related works}
The DRL-based algorithms have attracted extensive attention in the resource scheduling field. In the following, we present the related works from four aspects: Value-based DRL, Policy-based DRL, Distributed DRL, and Hybrid DRL.
 
Value-based DRL method: In \cite{liu2017enb}, a Q-learning algorithm with value-difference based exploration policy was proposed to solve the evolved NodeB (eNB) selection. In \cite{min2019learning}, a deep Q-leaning based offloading scheme was presented to optimize offloading policy according to the current battery level, the radio transmission rate and the harvested energy in the MEC system. In \cite{he2017integrated}, a deep Q network (DQN) approach was introduced to optimize the networking, caching, and computing resources in the  vehicular networks. Moreover, in \cite{chen2018optimized}, a double DQN based method with Q-function decomposition technique was applied to optimize stochastic computation offloading.

Policy-based DRL: In \cite{liu2018energy}, a deep deterministic policy gradient (DDPG) method was used to design the trajectory of UAVs by jointly considering the communications coverage, fairness, energy consumption and connectivity. Also, in \cite{jiang2020stacked}, DDPG was introduced to generate appropriate resource allocation decisions to satisfy the quality-of-service (QoS) requirements in an MEC system with vehicular applications. In \cite{feng2019cooperative}, the DRL was introduced to optimize the blockchain enhanced MEC system and an asynchronous advantage Actor-Critic (A3C)-based  computation offloading and resource allocation algorithm was applied to maximize the computation rate of MEC systems and the transaction throughput of blockchain system by jointly optimizing block size, block interval, offloading decision and power allocation.

Distributed DRL: In \cite{ye2019deep}, a distributed DQN was applied to search the optimal sub-band and power level for transmission, and each vehicle-to-vehicle link was considered as an agent and trained in a decentralized way. Then, in \cite{chen2018decentralized}, a decentralized offloading policy learned by DDPG for a multi-user MEC system was proposed, in which each DDPG was adopted to learn the efficient offloading policy interdependently at each mobile user.

Hybrid DRL method: In \cite{chen2019iraf}, the DRL with self-supervised learning was introduced to solve the complex resource allocation problem for a collaborative MEC network, in which the transitions are collected from Monte Carlo Tree Search in a self-play mode. In \cite{hua2019gan}, a generative adversarial network-powered deep distributional Q network (GAN-DDQN) was proposed to learn the action-value distribution driven by minimizing the discrepancy between the estimated and the target action-value distribution, which can be used to solve demand-aware resource allocation problem. 

However, none of the above contributions considered the application of distributed DRL in online resource scheduling for large-scale MEC systems, which can be seen as a complex MINLP problem. Traditional value-based and policy-based DRL cannot be applied in a large scale scenario. For example, the value-based DRL can only work well in a limited action space, thus it may not be suitable for large-scale offloading decision making environment. The output of policy-based DRL is continuous, so it is hard to output integer offloading variables and it is normally difficult to converge in dynamic environment \cite{lowe2017multi}. Distributed DRL has a great potential in addressing the above-mentioned issues by conducting distributed learning and dynamic decision in large-scale application scenarios \cite{chen2018decentralized}. However, the conventional contribution considered no global information of the whole system and just utilized the local information of each agent independently\cite{ye2019deep,chen2018decentralized}. 

In this paper, we propose a DIRS framework, by combining distributed decision making process and centralized training with the aid of ensemble learning and imitation learning, which has considerable performance gain over the conventional solutions.

%
%
%
%

\section{System model and problem formulation}
\subsection{System model}
\begin{figure}[htpb]
  \centering
	\includegraphics[width=8.8cm]{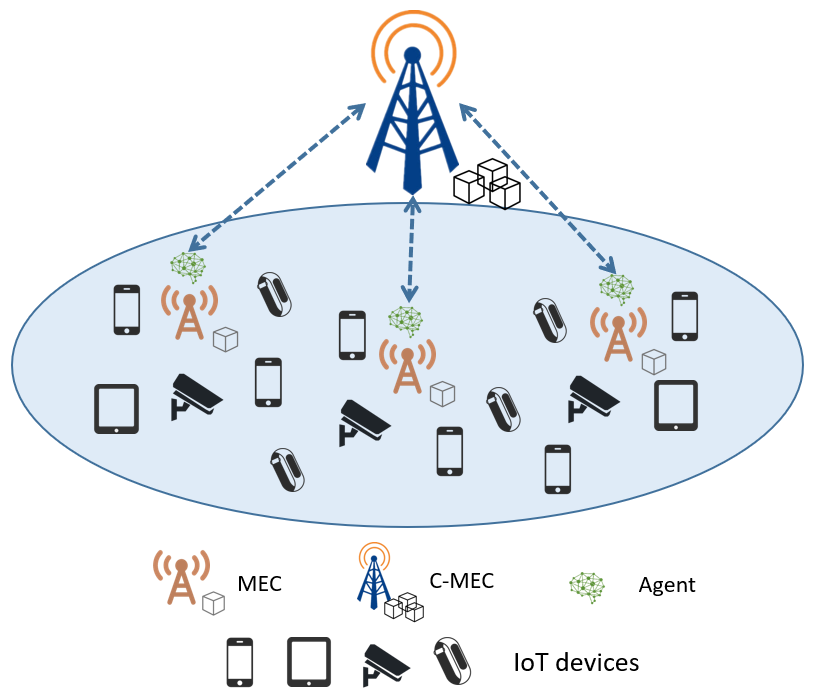}
	\caption{Proposed MEC system}
  \label{fig:fig1}
\end{figure}

Fig. \ref{fig:fig1} shows our proposed system with one C-MEC and $M$ normal MECs, denoted as the set of $\mathcal{M}=\{1,2,\ldots,M\}$. C-MEC is deployed at the macro base station, whereas the other edge servers are installed in each small base station. We assume there are $N$ IoTDs, denoted as the set of $\mathcal{N}=\{1,2,\ldots,N\}$. Each IoTD has a computation task to be executed, which can be either offloaded to the MECs or processed locally. 

We define the computing task in each IoTD as $U_{i}$, where 
$U_{i}=\left(F_{i}, D_{i}\right), \forall i \in \mathcal{N}$ \cite{wang2016joint}, $F_{i}$ denotes the total number of the CPU cycles and $D_{i}$ describes the data size transmitting to the MEC if offloading action is conducted. $D_{i}$ and $F_{i}$ can be obtained by using the approaches provided in \cite{yang2013framework}.

Then, the overall time consumption of completing  a task can be given by 
\begin{align}\label{eq:Shi5}
T_{ij} = T_{i j}^{T r}+ T_{i j}^{C}=\frac{D_{i}}{r_{i j}} + \frac{F_{i}}{f_{i j}}, \forall i \in \mathcal{N}, \forall j \in \mathcal{M}
\end{align}
where $T_{i j}^{T r}$ is time consumed for data offloading from the $i$-th IoTD to the $j$-th MEC,  $T_{i j}^{C}$ is execution time in an MEC server if the $i$-th IoTD offloading task to the $j$-th MEC server, $f_{i j}$ is the computation capacity of the $j$-th MEC allocating to the $i$-th IoTD and $j=0$ if IoTD executes the task locally. Also, $r_{i j}$ denotes the offloading data rate from the $i$-th IoTD to the $j$-th MEC, which can be given by
\begin{align}\label{eq:Shi11}
r_{i j}=B \log _{2}\left(1+\frac{p_{i j}^T h_{i j}}{\sigma^{2}}\right), \forall i \in \mathcal{N}, \forall j \in \mathcal{M}
\end{align}
where $B$ is the channel bandwidth, $\sigma^{2}$ is the noise spectral density and $h_{i j}$ is the channel gain which is given by
\begin{align}\label{eq:Shi10}
h_{i j}=\frac{\beta_{0}l_{ij}}{\left(X_{j}-x_{i}\right)^{2}+\left(Y_{j}-y_{i}\right)^{2}}, \forall i \in \mathcal{N}, \forall j \in \mathcal{M}
\end{align}
where $\beta_{0}$ denotes the channel power gain at the reference distance, $l_{ij}$ describes the small-scale fading factor, $\left(x_{i}, y_{i} \right)$ is the coordinate of the $i$-th IoTD, and $\left(X_{j}, Y_{j}\right)$ is the coordinate of the $j$-th MEC.


We consider a binary offloading strategy as
\begin{align}\label{eq:Shi1}
 a_{i j}=\{0,1\}, \forall i \in \mathcal{N}, \forall j \in \mathcal{M}
\end{align}
where $a_{i j}=1$, $j \neq 0$ denotes that the $i$-th IoTD decides to offload the task to the $j$-th MEC, while $a_{i j}=0$, $j\neq 0$ denotes that the $i$-th IoTD decides not to offload the task to the $j$-th MEC, and $a_{i j}=1$, $j=0$ denotes that the IoTD conducts the task locally. We assume that one IoTD can access to at most one edge server, which is formulated as follows:
\begin{align}\label{eq:Shi2}
 \sum_{j \in \mathcal{M'}} a_{i j} = 1, \forall i \in \mathcal{N}
\end{align}
where $\mathcal{M}^{\prime}=\{0,1,2,\ldots,M\}$ denotes the possible places at which the tasks can be executed and $j=0$ denotes that the task is conducted locally.

Also, assume that the computing resource of each MEC is constrained by
\begin{align}\label{eq:Shi8}
 \sum_{i=1}^{N} a_{i j} f_{i j} \leq F_{j, \ m a x }^{MEC}, \forall j \in \mathcal{M}
\end{align}
where $F_{j, \ m a x }^{M E C}$ is the computational capability of the $j$-th MEC.

Then, define $P_{i, \ m a x }^{IoTD}$ as the maximum transmission power that each IoTD can apply and then one has
\begin{align}\label{eq:Shi6}
  \sum_{j=1}^{M} a_{i j} p_{i j}^{T}+a_{i 0} p_{i}^{E} \leq P_{i, \ m a x }^{IoTD}
\end{align}
where $p_{i j}^{T}$ is the transmission power from the $i$-th IoTD to the $j$-th MEC server and $p_{i}^{E}$ is the execution power of the $i$-th IoTD which is given by 
$p_{i}^{E}=\kappa_{i}\left(f_{i 0}\right)^{v_{i}}, \forall i \in \mathcal{N}$, 
and $f_{i 0}$ is the local computing capacity and is a fixed value in this paper,
$\kappa_{i} \geq 0$ is the effective switched capacitance and $v_{i} \geq 1$ is the positive constant. To match the realistic measurements, we set $\kappa_{i}=10^{-27}$ and $v_{i}=3$ \cite{wang2016joint}.

\subsection{Problem Formulation}
For each IoTD, the time consumption is 
\begin{align}\label{12}
T_i=\sum_{j \in \mathcal{M}} a_{i j}  T_{ij} +a_{i 0} \frac{F_{i}}{f_{i 0}}.
\end{align}

Also, for each IoTD, the energy consumption is given by
\begin{align}\label{13}
E_i=\sum_{j \in \mathcal{M}} a_{i j} ( p_{ij}^T T_{ij}^{Tr} )+a_{i 0} \frac{F_{i}}{f_{i 0}} p_{i}^E. 
\end{align}

Then, define $\Phi_i$ as 
\begin{align}\label{14}
\Phi_i=  \phi_T T_i  +   \phi_E E_i.
\end{align}
where $\phi_T$ and $\phi_E$ are weighted coefficients. 

In this paper, we aim to jointly optimize the offloading selection, computing resource allocation, and power allocation to minimize the weighted sum of task latency and energy consumption of all tasks. Specifically, we formulate the optimization problem as follows:
\begin{displaymath}
\textit{P}0: \min _{\mathbf{a}, \mathbf{f}, \mathbf{p}} \sum_{i \in \mathcal{N}} \Phi_i
\end{displaymath}
\begin{equation}\label{eq:Shi15}
\text { s.t. } (4)-(7)
\end{equation}
where 
$\mathbf{a}=\left\{a_{i j}|i\in \mathcal{N}, j \in \mathcal{M}^{\prime}\right\}$, $\mathbf{f}=\left\{f_{i j}|i\in \mathcal{N},  j \in \mathcal{M}^{\prime}\right\}$, $\mathbf{p}=\left\{p_{i j}|i\in \mathcal{N},  j \in \mathcal{M}^{\prime}\right\}$ are vectors for offloading decisions, computing resource allocation and transmission power from each IoTD, respectively. One can see that this problem includes both integer and continuous variables. If IoTD conducts the task itself, the energy consumption can be expressed as $p_{i 0}=p_{i}^{E}$. Also assume that $\mathbf{h}=\left\{h_{ij} | i \in \mathcal{N},  j \in \mathcal{M}\right\}$ are time-varying input values, whereas other parameters are fixed.

\subsection{Problem transformation}
One can see that Problem $P$0 is an MINLP, which is very difficult to address in general. This problem becomes even more complex if it involves large-scale variables.   
To obtain the low complexity solution, we first decompose this problem into two sub-problems: 1) Offloading decision making sub-problem and 2) Resource allocation sub-problem. 

Firstly, we propose a distributed DRL framework to obtain optimal offloading decision $\mathbf{a}$ from the interaction between distributed agent and global environment. Once the offloading variable $\mathbf{a}$ is obtained, the resource allocation sub-problem can then be expressed as follows:
\begin{displaymath}
\begin{aligned}
\textit{P}1: &\min _{\mathbf{f}, \mathbf{p}} \sum_{i \in \mathcal{N}} \phi_{T}\left(\sum_{j \in \mathcal{M}} a_{i j}\left(\frac{D_{i}}{r_{i j}}+\frac{F_{i}}{f_{i j}}\right)+a_{i 0} \frac{F_{i}}{f_{i 0}}\right)+\\
&\phi_{E}\left(\sum_{j \in \mathcal{M}} a_{i j} p_{i j}^{T} \frac{D_{i}}{r_{i j}}+a_{i 0} \frac{F_{i}}{f_{i 0}} p_{i}^{E}\right)
\end{aligned}
\end{displaymath}
\begin{equation}\label{eq:Shi16}
\text { s.t. } (6)(7).
\end{equation}

Then, we can decouple the above problem into the following optimization to facilitate the distribute decision making in each MEC as 
\begin{displaymath}
\begin{aligned}
 &\min _{f_{j}, p_{j}} \sum_{i \in \mathcal{N}} \phi_{T}\left(a_{i j}\left(\frac{D_{i}}{r_{i j}}+\frac{F_{i}}{f_{i j}}\right)+a_{i 0} \frac{F_{i}}{f_{i 0}}\right)\\
&+\phi_{E}\left(a_{i j} p_{i j}^{T} \frac{D_{i}}{r_{i j}}+a_{i 0} \frac{F_{i}}{f_{i 0}} p_{i}^{E}\right), \forall j \in \mathcal{M}
\end{aligned}
\end{displaymath}
\begin{equation}\label{eq:Shi17}
\text {s.t. } (6)(7).
\end{equation}

The above problem only includes the continuous variables and therefore can be easily addressed using conventional convex optimization tools or heuristic optimization methods.


\section{Distributed DRL framework}
\begin{figure}[htpb]
	\centering
	\includegraphics[width=8.8cm]{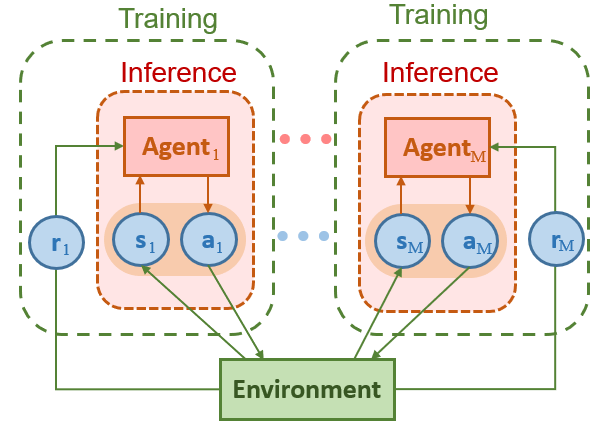}
	\caption{Distributed DRL paradigm with decentralized training and inference}
	\label{fig:fig2}
\end{figure}

\begin{figure}[htpb]
	\centering
	\includegraphics[width=8.8cm]{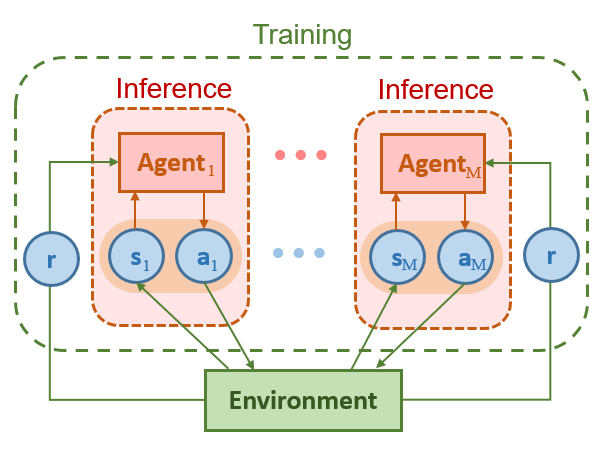}
	\caption{Distributed DRL paradigm with centralized training and decentralized inference}
	\label{fig:fig3}
\end{figure}
Before introducing our proposed framework, we first review two different modes of distributed DRL: (1) Decentralized training with decentralized inference (i.e., Fig. \ref{fig:fig2}) and (2) Centralized training with decentralized inference (i.e., Fig. \ref{fig:fig3}).

In the first mode, we have $M$ agents interacting with the environment in discrete decision epochs. The environment then generates a set of states $s_{t}=\left\{s_{1, t}, \ldots, s_{M, t}\right\}$ at the time slot $t$ and the agents carry out a set of actions $a_{t}=\left\{a_{1, t}, \ldots, a_{M, t}\right\}$  based on a set of policy $\pi_{t}=\left\{\pi_{1, t}, \ldots, \pi_{M, t}\right\}$ with parameters $\theta_{t}=\left\{\theta_{1,t}, \ldots,\theta_{M,t}\right\}$, which can be expressed as an unknown function mapping as:
\begin{equation}\label{eq:Shi18}
\pi_{t}=(s_{t} \rightarrow a_{t}).
\end{equation}

Then the environment produces a reward $r_{j, t}$ according to the action $a_{j, t}$ of the \textit{j}-th agent. For improving $a_{j, t}$, the \textit{j}-th agent updates the policy $\pi_{j, t}$ periodically until the current policy generates a locally optimal action $a_{j, t}^{*}$ to obtain the maximum local reward.

In the second mode, according to the actions $a_{t}=\left\{a_{1, t}, \ldots, a_{M, t}\right\}$ of all the agents, the environment produces a global reward $r_{t}$. For obtaining the maximum globally optimal reward $r_{t}$, each agent \textit{j} updates its policy $\pi_{j, t}$ periodically until all the policies $\pi_{t}=\left\{\pi_{1, t}, \ldots, \pi_{M, t}\right\}$ with parameters $\theta_{t}=\left\{\theta_{1,t}, \ldots,\theta_{M,t}\right\}$ achieving the globally optimal action $a_{t}^{*}$ to obtain the maximum global reward for all the agents in the training process. After training process, all the agents execute actions independently to make local inference. 

One can see that as the second mode considers the global reward in the training process, and therefore it potentially has better performance than the first mode, especially in large-scale application scenarios \cite{lowe2017multi}.
Therefore, in our problem, a distributed DRL is proposed by combing the centralized training and decentralized decision making process. Specifically, the C-MEC is responsible for the centralized training while each MEC can make its local decision in the inference stage. Next, we  give the proposed DIRS framework in details.


\section{Distributed intelligent resource scheduling (DIRS) framework}

In this section, we will introduce the proposed DIRS framework, which focuses on joint computation offloading decision and resource allocation in dynamic environment. Distributed model-free DRL is introduced to address the offloading decision making problem, as it is a goal-oriented method which can learn the optimal policy through the interaction between agent and environment. In a large-scale MEC system with multiple users, there are several challenges to be addressed as follows: (1) The policy is randomly distributed and the experience replay buffer is sparse at the beginning of the learning process, so that the interaction process is inefficient and the DRL framework is difficult to converge, especially in dynamic situations. (2) Because of the large number of users, the state space of the DRL is extremely large, which increases the difficulty of policy learning. (3) The action exploration is very challenging because of the complex optimization problem such that the DRL is difficult to explore the optimal action and the search is prone to trap into local minimum. These challenges prohibit the DRL from being directly applied in the real environment. To tackle the above-mentioned issues, we present a DRL-based DIRS framework with centralized training and decentralized inference in decision making stage. Next, we give a brief introduction of the proposed framework.

\subsection{The framework outline}
\begin{figure*}[htpb]
	\centering
	\includegraphics[width=18cm]{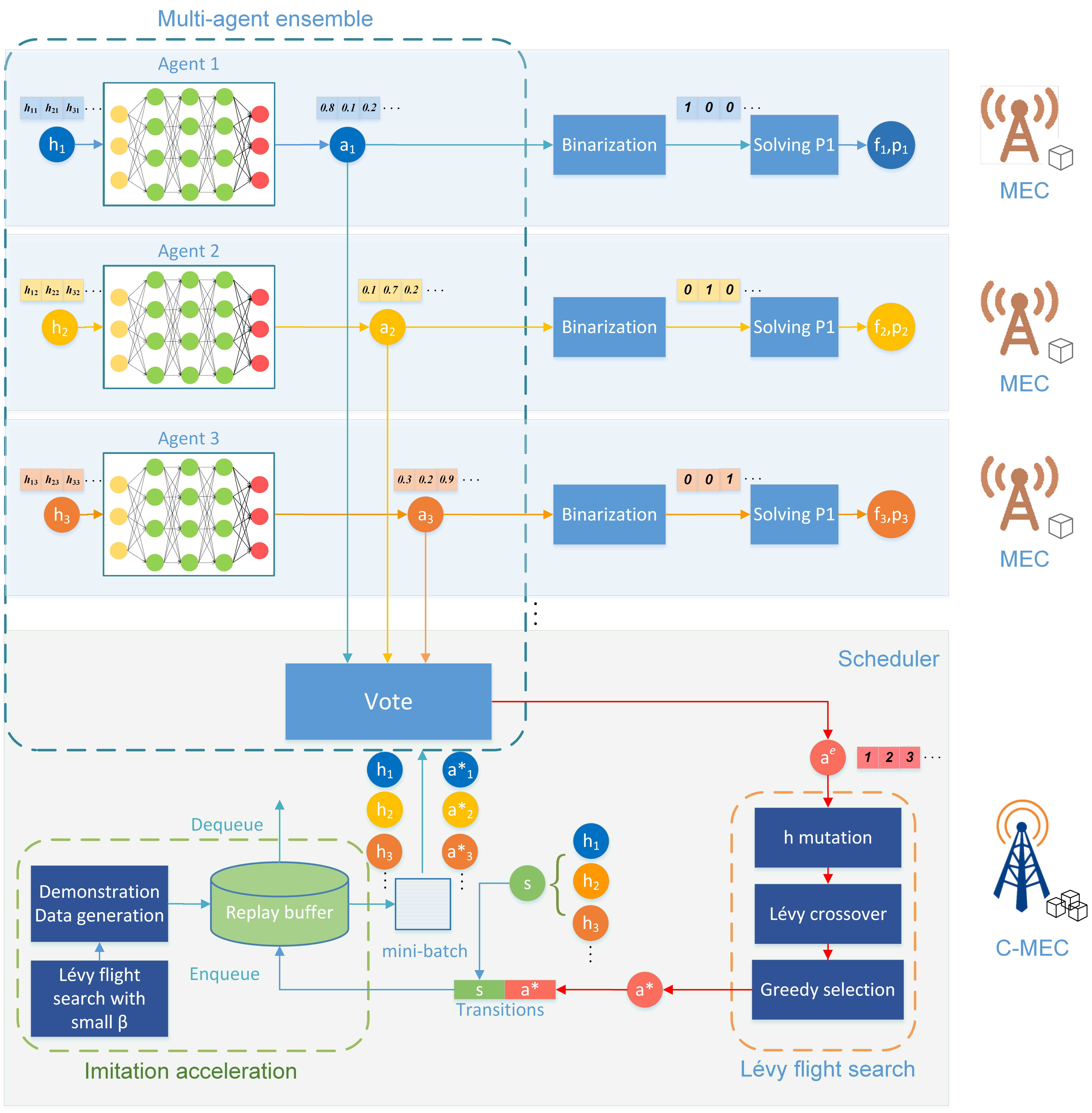}
	\caption{The DIRS framework}
	\label{fig:fig4}
\end{figure*}

The DIRS framework is illustrated in Fig. \ref{fig:fig4}, which includes local agents and a global scheduler. Specifically, the local agents are deployed on each MEC to conduct the independent decision making, and a global scheduler is deployed on the resource-abundant C-MEC to conduct the centralized training, which involves local information exchange of all the agents, generation of demonstration data by using imitation acceleration scheme and refinement of the actions with the aid of Lévy flight search. 
In online inference stage, each MEC only needs to perform some simple algebraic calculations with the help of local agent enhanced by DRL instead of solving the original optimization problem.

The workflow of the DIRS framework is presented in \textbf{Algorithm\enspace\ref{alg1}}. Firstly, we initialize parameters of all deep neural networks (DNNs) in $M$ agent with the parameters $\boldsymbol{\theta}^{0}=\left\{\theta_{1}^{0}, \ldots, \theta_{M}^{0}\right\}$ randomly generated and we also initialize an empty replay buffer $\mathcal{D}$. Then, we pre-train all the agents using demonstration data produced by the imitation acceleration scheme (i.e., \textbf{Algorithm\enspace\ref{alg4}}) and obtain the pre-trained parameters of all the agents $\boldsymbol{\theta}^{D}=\left\{\theta_{1}^{D}, \ldots, \theta_{M}^{D}\right\}$, and then keep all the demonstration data in the replay buffer $\mathcal{D}$. Next, the online training stage and online distributed inference stage are executed. Particularly, the online distributed inference stage is performed continuously, where each agent \textit{j} generates an offloading action $a_{j}=\left\{a_{1 j}, \ldots, a_{N j}\right\}$ for $N$ IoTDs according to the channel state information $h_{j}=\left\{h_{1 j}, \ldots, h_{N j}\right\}$, and obtains $p_{j}=\left\{p_{1 j}, \ldots, p_{N j}\right\}$ and $f_{j}=\left\{f_{1 j}, \ldots, f_{N j}\right\}$ by solving Problem (\ref{eq:Shi17}) independently. The online training stage is performed at every interval $\phi$ by multi-agent ensemble algorithm (i.e., \textbf{Algorithm\enspace\ref{alg2}}) and Lévy flight search algorithm (i.e., \textbf{Algorithm\enspace\ref{alg3}}), and then the learned parameters of all $M$ agents $\boldsymbol{\theta}=\left\{\theta_{1}, \ldots, \theta_{M}\right\}$ are updated. These two stages are alternately performed and the offloading policies of all the agents can be gradually improved in the iteration process.

\begin{algorithm}
	\caption{DIRS framework}
	\label{alg1}
	\begin{algorithmic}[1]
		\REQUIRE   $h_{j, t}, T_{D R L}, \phi$.
		\ENSURE $a_{j, t}, p_{j, t}, f_{j, t}$.
		\STATE{Initialize $M$ agents with policies randomly parameterized by $\boldsymbol{\theta^{0}}$.}
		\STATE{Initialize an empty replay buffer $\mathcal{D}$.}\\
		$\bf{Offline\;pre-training\;stage}$\\
		\STATE{Train all agents using demonstration data by \textbf{Algorithm\enspace\ref{alg4}}, and obtain the demonstrated parameters of all agents $\boldsymbol{\theta^{D}}$.}\\
		\STATE{Keep all demonstration data in the replay buffer $\mathcal{D}$.}\\
		$\bf{Online\;decentralized \;inference\; stage}$\\
		\WHILE{$t<T_{D R L}$}
		\FOR{each agent \textit{j}}
		\STATE{Obtain the channel state information $h_{j, t}$ according to the environment.}
		\STATE{Generate the offloading action $a_{j, t}=\pi\left(h_{j, t} | \theta_{j, t}\right)$ independently.}
		\STATE{Obtain $p_{j, t}$ and $f_{j,t}$ by solving Problem (\ref{eq:Shi17}) independently.}
		\ENDFOR\\
		$\bf{Online\;training\; stage}$\\
		\IF{$t \bmod \phi=0$}
		\STATE{Train all agents using \textbf{Algorithm\enspace\ref{alg2}} and \textbf{Algorithm\enspace\ref{alg3}}}.\\
		\ENDIF
		\ENDWHILE
	\end{algorithmic}
\end{algorithm}

In general, the DIRS includes three work stages: (1) Offline pre-training stage is applied to accelerate the DRL training for the large-scale application scenarios; (2) Online training stage is introduced to track
the variations of the real scenarios in dynamic environments; (3) Online decentralized inference stage is presented to make real-time decisions. Moreover, there are three key improvements of the DIRS framework compared to traditional DRL: (1) An imitation acceleration scheme is presented to generate demonstration data and initialize the parameters of distributed agents rather than initializing them randomly to accelerate the learning speed of the distributed DRL (i.e., Subsection-V-B). (2) Multi-agent ensemble algorithm is proposed in Subsection-V-C for large state space partition and decision consolidation, in which the original channel state information $\mathbf{h}$ is regarded as the current state $s$ of DRL and it is divided into smaller subsets $\left\{h_{1}, h_{2}, \ldots, h_{M}\right\}$ according to the ownership of MECs. 
Finally, the maximum vote approach is applied to integrate the results of all distributed agents, and obtain the ensemble offloading decision $\mathbf{a}^{\boldsymbol{e}}$ according to the maximum vote. This method can realize dimensionality reduction for the DNN in each agent and simplify the policy learning in each sub-state space. (3) A novel Lévy flight search is introduced in Subsection-V-D for action refinement, which can enhance the action exploration of DRL. Then the optimal offloading action $\mathbf{a}^{\boldsymbol{*}}$ is achieved by maximizing the reward which is cached into the replay buffer $\mathcal{D}$. One can see that the DIRS framework is a model free DRL which can provide distributed decision making and resource allocation without solving the original MINLP problem. We describe the implementation details of each module in the following subsections.

\subsection{Imitation acceleration scheme}
In large-scale scenarios, distributed DRL typically requires to learn a huge amount of data before they reach reasonable performance, which is very time-consuming by trial and error. This is the major drawback of the DRL to solve the large-scale optimization problem. Recently, imitation learning has been shown to help address this difficult exploration problems in DRL \cite{hester2018deep}.

Imitation learning focuses on imitating human learning or expert demonstration for controlling the behaviour of the agent, which can help DRL reduce the time required to learn by an agent to a great extent through reducing the number of trials \cite{liu2020analyzing}. In DRL, imitation learning can help an agent to achieve better performance in complex environment by pre-training it with the demonstration data.


In the proposed DRL, we propose an imitation acceleration scheme 
combined with DRL and imitation learning. The imitation acceleration algorithm leverages relatively a small amount of demonstration data to pre-train the agents in our framework, which can significantly accelerate the learning process of the distributed DRL. The details of the imitation acceleration scheme are described as follows.

Firstly, we collect demonstration data by leveraging the optimization algorithm for solving the problem in Eq. (\ref{eq:Shi15}). In general, the algorithm can be divided into three categories: (1) If the action space is small, we can use the exhaustive search approach to obtain the optimal decision. (2) If the action space is medium, we can use some mixed integer programming solver (e.g., CPLEX). (3) If the action space is large, we can use some global heuristic algorithms to obtain suboptimal decisions \cite{yu2020intelligent}. In our study, the demonstration data is generated from the Lévy flight search with small value of  parameter $\beta$ (to be introduced in Subsection-V-D) which is suitable for solving large-scale MINLP problems and can possibly achieve globally optimal solutions \cite{emary2019impact}. Then the channel state information as well as its optimal offloading actions solved by the Lévy flight search are stored into the replay buffer $\mathcal{D}$.

Secondly, we define a novel demonstration loss function, which is a combination of two losses:
\begin{equation}\label{eq:Shi29}
L_{1}(\theta)=L_{D}(\theta)+\lambda_{1} L_{L 2}(\theta)
\end{equation}
where  $L_{D}(\theta)=-\frac{1}{K} \sum_{k \in \mathcal{K}}\left(\left(a_{k}^{*}\right)^{T} \log \left(\pi\left(h_{k} | \theta\right)\right)+\left(1-a_{k}^{*}\right)^{T}\right.$
$\log \left(1-\pi\left(h_{k} | \theta\right)\right)$
is the demonstration data loss which is the cross-entropy for demonstration data and $K$ is the number of demonstration data in the batch,  whereas $L_{L 2}(\theta)=\|\theta\|^{2}$ is the L2-norm of $\theta$, which can increase the generalization of DNN in each agent.

Third, we initialize the pre-training process of all the agents solely on the demonstration data before starting any interaction with the environment.

Finally, once the offline pre-training phase is complete, all the agents start to act in the environment. Then the online training stage and distributed inference stage are alternately executed. We describe the whole process of \textbf{Algorithm\enspace\ref{alg4}} as follows.
\begin{algorithm}
	\caption{Imitation acceleration algorithm}
	\label{alg4}
	\begin{algorithmic}[1]
		\REQUIRE $T_{D}$.
		\ENSURE pre-trained parameters $\boldsymbol{\theta}^{D}$.
		\STATE{Generate demonstration data and record them in the replay buffer $\mathcal{D}$.}
		\WHILE{$t \textless T_{D}$}
		\STATE{ Sample a minibatch of $K$ transitions from replay buffer $\mathcal{D}$.}
		\STATE{Train all agent and update the offloading policy $\pi_{t}$ with parameters $\boldsymbol{\theta}^{D}$ using the demonstration loss in Eq. (\ref{eq:Shi29}).}
		\ENDWHILE
	\end{algorithmic}
\end{algorithm}

\subsection{Multi-agent ensemble based distributed DRL algorithm }
In the proposed large-scale MEC systems with a large number of IoTDs, the traditional DRL suffers from the challenge that the state space is also extremely large, and therefore it is difficult to train a steady policy because of the partial observability of the large state space. To address this, we introduce ensemble learning to enhance the distributed DRL as follows.

Ensemble learning is a machine learning method that generates and combines multiple inducers to solve the complex optimization problem. The intuitive explanation behind ensemble learning stems from human nature and the tendency to gather different opinions and combine them to make a complicated decision \cite{krawczyk2017ensemble}. There are many advantages to introduce ensemble learning into our DIRS framework as follow.

(1) Global optimization: a single agent that conducts local inference may get stuck in a local optimum. By combining several agents, ensemble methods can decrease the risk of obtaining a locally minimal solution.

(2) Dimensionality reduction: each agent can be constructed and trained using a selected subset of all the features (i.e., its own channel gains), which can reduce the impact of the curse of dimensionality by reducing the state space for each agent.

(3) Output manipulation: the original offloading decision can be seen as a multi-class classification problem for multiple MECs, but each agent is a binary classifier. By applying ensemble learning, many binary classifiers can be combined into a multi-class classifier, which can simplify the decision process of each agent.

To this end, we propose a multi-agent ensemble algorithm to support the distributed DRL, in which the basic components are redefined as follows.

(1) State space: The overall channel state information is  $\mathbf{h}=\left\{h_{i j}\right\}, \forall i \in \mathcal{N}, \forall j \in \mathcal{M}$. We also assume that each agent in each MEC can only obtain its own related channel state information, and therefore the state of agent \textit{j} is denoted as $h_{j}=\left\{h_{1 j}, \ldots, h_{N j}\right\}$.

(2) Action space: We define two kinds of action, i.e., global action and local action for each agent. Specifically, global action is applied in the training stage, whereas local action is generated in each agent for the distributed decision making. The global action is defined as $\mathbf{a^e}=\left\{a_{i}\right\}, \forall i \in \mathcal{N}$, where $a_{i}=0$  means the $i$-th IoTD decides to execute the task itself, whereas $a_{i}=j$ means that the $i$-th IoTD decides to offload the task to the $j$-th MEC, where $j \in \mathcal{M}$. The local decision in the $j$-th agent is defined as $a_{j}=\left\{a_{1 j}, \ldots, a_{N j}\right\}$, where $a_{i j}=1$ means that the $i$-th IoTD decides to offload the task to the $j$-th MEC, while $a_{i j}=0$ means the \textit{i}-th IoTD decides not to offload the task to the \textit{j}-th MEC. One can see that each agent can make distributed decision according to its own local information. 

(3) Reward: The global reward is defined as the reciprocal of the objective function in Eq. (\ref{eq:Shi15}). 


(4) Policy: The policy $\pi_{j}\left(h_{j} | \theta_{j}\right)$ of each agent \textit{j} is implemented by applying a DNN, where $\theta_{j}$ is the parameters of the DNN at the \textit{j}-th agent.


(5) Vote: The output of each agent can be combined by voting solutions. The final solution is chosen based on the highest value of the decisions of all the agents. Here we give an example to illustrate the proposed voting process. Suppose there are three MECs and the outputs of three agents for the $i$-th IoTD are $a_{i1}$=0.7,  $a_{i2}$=0.8 and $a_{i3}$=0.2, respectively. Since the second value $a_{i2}$ is the highest number, we can have the global ensemble action of the $i$-th IoTD $a_{i }^{e}$ as 2, which means this IoTD will offload the task to the second MEC. If all the values of the  $i$-th IoTD are lower than 0.5, then we set the corresponding ensemble action $a_{i }^{e}$ of the $i$-th IoTD as 0.

(6) Prioritized experience replay: The scheduler in C-MEC refines the global action by applying Lévy flight search (to be introduced in Subsection-V-D) and adds the self-generated data to its replay buffer $\mathcal{D}$. Data will be added to the replay buffer until it is full, and then the agent starts to over-write old data in this buffer during the online training stage. 
Prioritization batch is also used in the replay buffer. The probability of sampling transitions \textit{i} is defined as 
\begin{equation}\label{eq:Shi30}
P_{i}=\frac{w_{i}^{\tau}}{\sum_{k \in \mathcal{K}} w_{k}^{\tau}}
\end{equation}
where $w_{i}=\left|\Delta \delta_{t}\right|+\epsilon$ and $\epsilon$ is small positive constant that guarantees that all the transitions can be sampled. Two different $\epsilon_{A}$ and $\epsilon_{D}$ can be applied to control the relative sampling of the agent generated data versus the demonstration data. $\Delta \delta_{t}$  is the average loss function variation of all the agents. The exponent $\mathcal{\tau}$ determines the intensity of prioritization.

(7) Loss function: The loss function of the DNN in the agent is proposed as follows:

\begin{equation}\label{eq:Shi19}
L_{2}(\theta)=L_{1}(\theta)+\lambda_{2} L_{A}(\theta)
\end{equation}
where $L_{1}(\theta)$ is the demonstration loss function defined in Eq. (\ref{eq:Shi29}) and $L_{A}(\theta)$ is the  agent generated data loss which is given by
\begin{equation}\label{eq:ShiX2}
\begin{aligned}
L_{A}(\theta)=-\frac{1}{S} \sum_{k \in S}\left({\left(a_{k}^{*}\right)^{T} \log \left(\pi\left(h_{k} | \theta\right)\right)}\right. \\
\left. {+\left(1-a_{k}^{*}\right)^{T} \log \left(1-\pi\left(h_{k} | \theta\right)\right)}\right)\\
\end{aligned}
\end{equation}
where $S$ is the number of agent generated transitions in the batch.


When we initialize the environment and obtain the state, the offline training stage is performed, in which each agent \textit{j} generates the offloading action $a_{j, t}$ according to the policy $\pi_{j, t}\left(h_{j, t} | \theta_{j, t}\right)$ at the time slot $t$. Then, we obtain the global offloading decision $a_{t}^{e}$ according to the maximum vote of all the agents. To increase efficiency of action exploration, we search the best $a_{t}^{*}$ by applying \textbf{Algorithm\enspace\ref{alg3}}, and then we append the state-action pairs $\left\{h_{t}, a_{t}^{*}\right\}$ to the replay buffer $\mathcal{D}$ as training samples of all the agents. Next, we sample a batch of transitions by applying priority strategy discussed before, and then we train all the agents using the Adam algorithm\cite{kingma2014adam} and update all the offloading policies $\pi_{t}=\left\{\pi_{1, t}, \ldots, \pi_{M, t}\right\}$ by minimizing the loss function defined in Eq. (\ref{eq:Shi19}).  We describe the whole process of \textbf{Algorithm\enspace\ref{alg2}} as follows.
\begin{algorithm}
	\caption{Multi-agent ensemble algorithm}
	\label{alg2}
	\begin{algorithmic}[1]
		\REQUIRE  $h_{t}, a_{t}, \mathcal{D}$.
		\ENSURE $\theta_{t}$.
		\STATE{Integrate the global $a_{t}^{e}$ by voting form all agents.}
		\STATE{Find the best $a_{t}^{*}$ by \textbf{Algorithm\enspace\ref{alg3}}.\\
			\STATE{Append the state-action pair $\left\{h_{t}, a_{t}^{*}\right\}$ to the replay buffer $\mathcal{D}$. }
			\STATE{Sample a random minibatch of transitions by priority strategy using Eq. (\ref{eq:Shi30}) from replay buffer $\mathcal{D}$.}
			\STATE{Feed these transitions to all agents.}
			\FOR{each agent \textit{j}}
			\STATE{Update agent parameters $\theta_{j, t}$ by minimizing the loss function in Eq. (\ref{eq:Shi19}) in distributed way.}
			\ENDFOR }
		\end{algorithmic}
	\end{algorithm}

\subsection{Lévy flight search}
Action exploration is an important part in DRL. Traditional DRL normally uses random process for action exploration (e.g., $\epsilon$-greedy), which is blind and inefficient in large action space\cite{lowe2017multi}. Some local search methods are also applied for action exploration which can achieve the suboptimal offloading policy $\pi$ \cite{huang2019deep,dulac2015deep}. However, these local search methods are easily stuck in local optimum and the right offloading policy cannot be guaranteed, especially in large-scale MEC systems \cite{huang2019deep}. To address this problem, we introduce a novel Lévy flight search, which can jump out of local optimum with high efficiency, to find the best offloading action $a_{t}^{*}$ according to $a_{t}^{e}$ in DRL. After Lévy flight search, the newly generated state-action pairs $\left\{s_{t}, a_{t}^{*}\right\}$ at time slot $t$ are appended to the replay buffer as training transitions of all the agents.

In recent years, Lévy flight as a heuristic algorithms has been applied to various areas  \cite{jiang2018electrical,emary2019impact}. One remarkable merit of Lévy flight is that it can explore the search space more efficiently than standard Gaussian process considering its heavy-tailed effect \cite{jiang2018using}.  Specifically, Lévy flight mainly consists of small steps for local search and occasionally large steps or long-distance jumps for escaping from the local optimum. Such small steps can accelerate the convergence speed while long jumps can avoid the problem of the algorithm’s premature, especially in a large-scale action space.

Lévy flight is a random process whose step distance is drawn from Lévy distribution. Mathematically speaking, the Lévy distribution has a simplified form as follows:
\begin{equation}\label{eq:Shi21}
\text { Lévy }(s)=\frac{1}{\pi} \int_{0}^{\infty} \exp \left(-\alpha|k|^{\beta}\right) \cos (k s) d k
\end{equation}
where $1<\beta \leq 2$ is a variation factor. This integral turns into the Cauchy distribution when $\beta=1$, while it turns into the Gaussian distribution when $\beta=2$. Also, $\alpha>0$ is the scaling factor and $s$ is the step length of Lévy flight, which can be calculated as follows \cite{mantegna1994fast}:

\begin{equation}\label{eq:Shi22}
d=\frac{u}{|v|^{\frac{1}{\beta}}}
\end{equation}
where $u$ and $v$ are drawn from Gaussian distribution, which can be calculated by

\begin{equation}\label{eq:Shi23}
u \sim N\left(0, \sigma_{u}^{2}\right), \quad v \sim N(0,1)
\end{equation}
where $\sigma_{u}$ can be calculated by

\begin{equation}\label{eq:Shi24}
\sigma_{u}=\left\{\frac{\Gamma(1+\beta) \cdot \sin (\pi \cdot \beta / 2)}{\Gamma[(1+\beta) / 2] \cdot \beta \cdot 2^{(1+\beta) / 2}}\right\}^{\frac{1}{\beta}}
\end{equation}
where $\Gamma$ is the standard Gamma function.

Traditional Lévy flight is used to generate a new solution in the heuristic search as follows:
\begin{equation}\label{eq:ShiX}
x_{i}(G)=x_{i}(G-1)+\eta d_{i}
\end{equation}
where $x_{i}(G)$ is the solution at iteration $G$, $d_{i}$ is a Lévy flight step, and $\eta$ is a scale weight.

However, there are two disadvantages that avoid Lévy flight from being directly applied to our distributed DRL algorithm. Firstly, the search step $d_{i}$ output from Lévy flight is a continuous real number, but the offloading decision $a_{i}$ in our problem is an integer value. Secondly, it does not take advantage of the channel state information in the search process. To tackle these issues, we propose a novel Lévy flight search, in which the solution can be represented as:
\begin{equation}\label{eq:Shi25X}
x=[a_1,\ldots,a_{N},f_1,\ldots,f_{N},p_1,\ldots,p_{N}]
\end{equation}
where $N$ is the number of IoTDs, $a_i$, $f_i$ and $p_i$ denote the offloading decision, allocated computation resource and transmission power of the $i$-th IoTD, respectively.

The Lévy flight search consists of three operations: h mutation, Lévy selection and greedy selection. We introduce each of them as follows.

(1) h mutation: channel state information $\mathbf{h}$ provides the prior information to create a mutant vector $a_{i}^{m}(G)$. The h mutation is applied by comparing the normalized $h_{i j}$ with a Lévy flight step, which represents that the IoTD that offloads the task to the MEC with higher $h_{i j}$ has a higher stability. Thus, the h mutation for the integer part of the solution can be represented as follows:

\begin{equation}\label{eq:Shi25}
a_{i}^{m}(G)=\left\{\begin{aligned}
&randm & \text { if } \gamma  d_{i}>\frac{h_{i, a_{i}}}{\sum_{j \in \mathcal{M}} h_{i j}} \\
&a_{i}(G-1) & \text { otherwise }\\
\end{aligned}\right.
\end{equation}
where  $randm$ $\in \mathcal{M}^{\prime}$ is a randomly generated integer to ensure that the \textit{i}-th IoTD will offload the task to an MEC or execute the task itself, $\gamma$ is a decreased weight calculated as follows:

\begin{equation}\label{eq:Shi27}
\gamma=2-2 G / G_{max}
\end{equation}
where $G_{max}$ is the maximum iteration number.

The rest part of the solution (e.g., $f_{i}^{m}(G)$ and $p_{i}^{m}(G)$) is generated by Eq. (\ref{eq:ShiX}).

2) Lévy crossover: the Lévy crossover is carried out to produce a candidate vector $x_{i}^{c}(G)$ by combining the mutant vector $x_{i}^{m}(G)$ and a target vector $x_{i}(G-1)$, which is achieved by comparing the weighted Lévy flight step with a threshold $t h$. Thus, the Lévy crossover is represented as follows:

\begin{equation}\label{eq:Shi26}
x_{i}^{c}(G)=\left\{\begin{aligned}
&x_{i}^{m}(G) & \text { if } \gamma d_{i}>t h \\
&x_{i}(G-1) & \text { otherwise }\\
\end{aligned}\right.
\end{equation}

3) Greedy selection: The selection operator determines whether the candidate vector $x_{i}^{c}(G)$  or the target vector $x_{i}(G-1)$ survives into the next iteration. Greedy selection is used to select the vector with the better fitness as follows:

\begin{equation}\label{eq:Shi28}
x_{i}(G)=\left\{\begin{aligned}
&x_{i}^{c}(G) \qquad \text { if } f\left(x_{i}^{c}\right)<f\left(x_{i}(G-1)\right) \\
&x_{i}(G-1) \qquad \text { otherwise }\\
\end{aligned}\right.
\end{equation}
where $f(\cdot)$ denotes the objective function in Eq. (\ref{eq:Shi15}). 

In a word, the Lévy flight search employs the channel state information to guide the mutation of solution, and introduces Lévy steps to avoid solution trapping into the local optimum, so it is an efficient action exploration method. The detailed description of the Lévy flight search algorithm is provided in \textbf{Algorithm\enspace\ref{alg3}}.

\begin{algorithm}
	\caption{Lévy flight search algorithm}
	\label{alg3}
	\begin{algorithmic}[1]
		\REQUIRE $a_{t}^{e}, \beta, t h$
		\ENSURE  $a_{t}^{*}$
		\STATE{Initialize $x(0)$ with ensemble offloading decision $a_{t}^{e}$, and random computation resource and energy resource allocation.}
		\WHILE{$G \leq G_{max}$}
		\STATE{Generate Lévy search step $d_{i}$ by Eq. (\ref{eq:Shi22})-Eq. (\ref{eq:Shi24}).}
		\STATE{Generate a mutant vector $x_i^m (G)$ with $a_i^m (G)$ by Eq. (\ref{eq:Shi25})-Eq. (\ref{eq:Shi27}), and $f_i^m (G)$ and $p_i^m (G)$ by Eq. (\ref{eq:ShiX}) in h mutation.}
		\STATE{Obtain a candidate vector $x_i^c (G)$ by Lévy crossover in Eq. (\ref{eq:Shi26}).}
		\STATE{Calculate the fitness of $x_i^c (G)$ and $x_{i}(G-1)$ by solving the objective function in Eq. (\ref{eq:Shi15}).}
		\STATE{Select the subsequent solution $x_{i}(G)$ by greedy selection in Eq. (\ref{eq:Shi28}).}
		\STATE{Update $\gamma$ according to Eq. (\ref{eq:Shi27}).}
		\ENDWHILE
	\end{algorithmic}
\end{algorithm}

In addition, variation factor $\beta$ is the key parameter of Lévy flight to balance global and local search. If $\beta$ value is large, the step size of
Lévy flight will be restricted in a small search range, which can really focus on the local search and occasionally global search. Therefore, larger $\beta$ value implies the faster convergence speed and slightly lower solution accuracy. If the $\beta$ value is small, the search length of walking distance of Lévy flight is long so that the global search can be enhanced and the optimal solution is achieved with slow convergence speed. For these above reasons, we use Lévy flight search with large $\beta$ value in online training stage (i.e., Ensemble learning at\textbf{ Algorithm 3}) and apply Lévy flight search with small $\beta$ value in offline pre-training stage (i.e. Imitation acceleration at \textbf{Algorithm 2}).

\section{Simulation results and numerical analysis}
\subsection{Simulation environment settings}
We first present experimental settings of the MEC scenario in Table \ref{tab:table1}. The parameters of the DIRS framework are chosen as follows: $T_{DRL}$=3000, replay buffer size=1024, minibatch size=256 and training interval $\phi$=10. We use a 4-layer fully-connected DNN in the agent, which includes 30, 80, 60 and 30 neurons in each layer, respectively. The parameters of the imitation acceleration scheme are chosen as follows: $T_{D}$=1000, $\lambda_{1}$=$10^{-4}$. The parameters of
the multi-agent ensemble algorithm are chosen as follows: $\lambda_{2}$=0.5, $\epsilon_{A}$=0.08 and $\epsilon_{D}$= 0.02. The parameters of the Lévy flight search are chosen as follows: $\beta$=1.5 and $th$=0.5. We assume there are two MEC servers and 30 IoTDs randomly distributed in the squared area with size
50m$\times$50m. Next, we present two different evaluations to verify the performance of the DIRS framework.

\begin{table}[]
	\centering\makegapedcells
	\caption{Simulation parameters}
	\label{tab:table1}
	\begin{tabular}{|l|l|}
		\hline
		$\bf{Parameters}$  & $\bf{Assumptions}$ \\ \hline
		Data size of task  ${D_i}$  & 100kB \\ \hline
		Required CPU cycles of task  ${F_i}$  & $10^9$ cycles/s \\ \hline
		Bandwidth $B$  & 1MHz \\ \hline
		Local Computational Capability $f_{i0}$  & $10^9$ cycles/s \\ \hline
		Remote Computational Capability $F_{max}^{MEC}$  & $50 \cdot 10^9$ cycles/s \\ \hline
		Maximum transmission power $P_{max }^{IoTD}$ & 1.5W \\ \hline
		Noise Spectral Density  $\sigma^{2}$ & $10^{-12}$W/Hz \\ \hline
	\end{tabular}
\end{table}

\subsection{Centralized training performance evaluation}

Imitation learning is used as a pre-training tool in our DIRS framework, and the amount of demonstration data will influence the performance of pre-training. We compare the performance of imitation learning with various quantities of demonstration data using different loss functions in TABLE \ref{tab:table3}. 
It can be stated that, with the growth of the demonstration data quantity, the training accuracy and testing accuracy first increases and then stabilize when the demonstration data quantity is above 256. This is because increasing demonstration data can allow the DNNs to learn more information from the problem. However, the DNNs may not improve its learning performance significantly when demonstration data quantity is above 256. For this reason, in
the follow simulations, the demonstration data quantity is set to 256. It also can be observed that the proposed loss function $L_{1}(\theta)$ achieves better training accuracy and testing accuracy than $L_{D}(\theta)$.
The reason behind this is that the $L_2$ regularized term is added in $L_{1}(\theta)$ and then the regularized term ensures the generalization of DNN, which leads to a higher learning accuracy, especially for the testing process.

\begin{table}[]
		\centering\makegapedcells
	\caption{The performance comparison of imitation learning.}	
	\label{tab:table3}
	\begin{tabular}{|p{40pt}<{\centering}|p{38pt}<{\centering}|p{38pt}<{\centering}|p{38pt}<{\centering}|p{38pt}<{\centering}|}
	\hline
	\multirow{2}{*}{Data quantity} & \multicolumn{2}{l|}{$L_{1}(\theta)$} & \multicolumn{2}{l|}{$L_{D}(\theta)$} \\ \cline{2-5} 
	&   Training accuracy    &      Testing accuracy     &   Training accuracy    &     Testing accuracy     \\ \hline
 64 	&   90.8\%      &        90.1\%   &    90.3\%       &     85.5\%      \\ \hline
 128 	&    92.4\%     &      91.8\%     &     92.2\%      &    89.8\%       \\ \hline
 256 	&   96.7\%      &    96.3\%       &      96.6\%     &     93.4\%      \\ \hline
 512	&   96.7\%       &      96.5\%     &      96.6\%     &     94.3\%      \\ \hline
\end{tabular}
\end{table}

\begin{figure}[htpb]
	\centering
	\includegraphics[width=8.8cm]{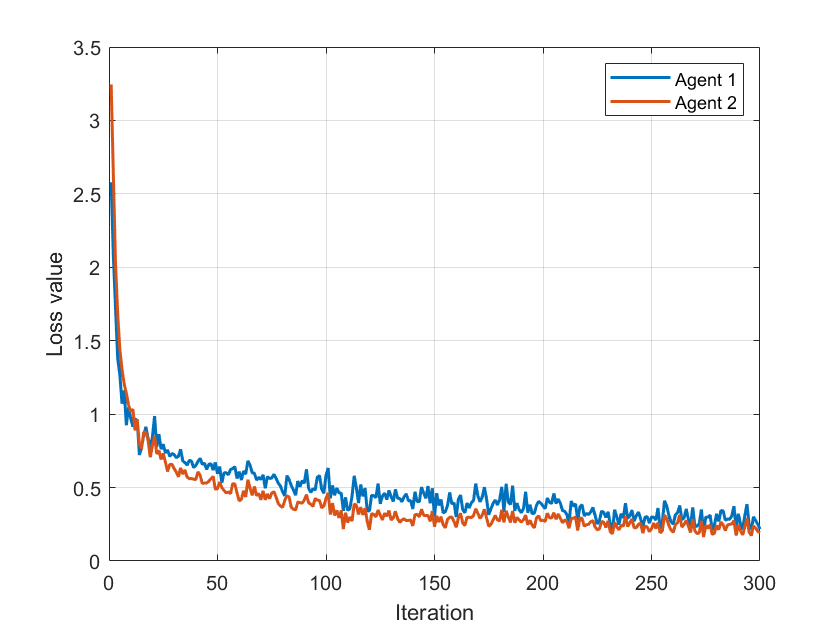}
	\caption{Comparison of loss values for Agent1 and Agent2}
	\label{fig:fig5}
\end{figure}

Multi-agent ensemble learning assisted DRL is the core part of DIRS framework, which uses multi-agent learning and decision consolidation. Fig. \ref{fig:fig5} shows the convergence curves of all the agents in different MECs.  One can see that the loss values of agent 1 and agent 2 all converge to 0.2 after around 300 iterations, which means all the agents in our DIRS framework can work properly and the over-fitting does not happen. 

\begin{figure}[htpb]
 	\centering
 	\includegraphics[width=8.8cm]{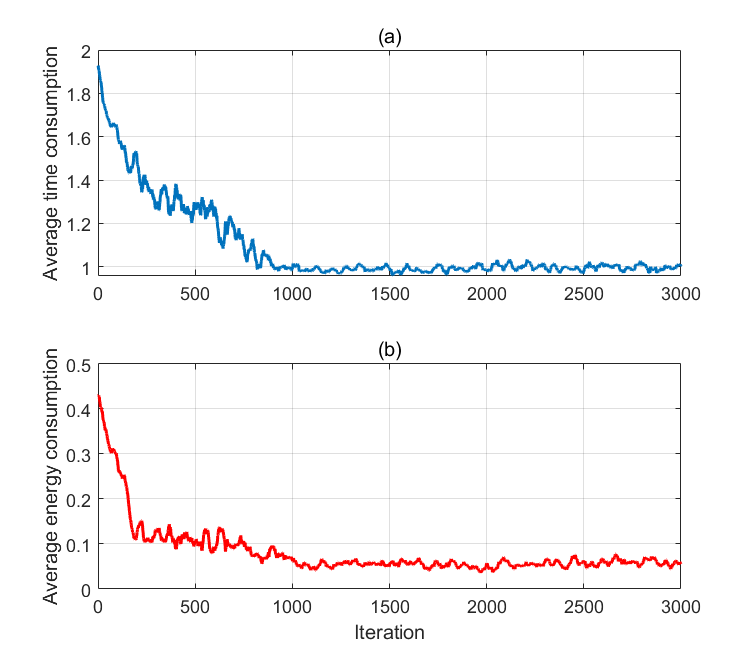}
 	\caption{Average time and energy consumption}
 	\label{fig:fig6}
 \end{figure}

Fig. \ref{fig:fig6} describes the performance of the average time consumption and energy consumption during the centralized learning process. From Fig. \ref{fig:fig6} (a), one can see that the average time consumption decreases dramatically in the early 900 iterations and then reaches a stable status. From Fig. \ref{fig:fig6} (b), one can see that the decrease speed of average energy consumption begins to reduce when the iteration number reaches 1100. 
As shown in Fig. \ref{fig:fig6}, our proposed framework can achieve considerable performance efficiently through the interaction between distributed agents and the global environment.

\begin{figure}[htpb]
	\centering
	\includegraphics[width=8.8cm]{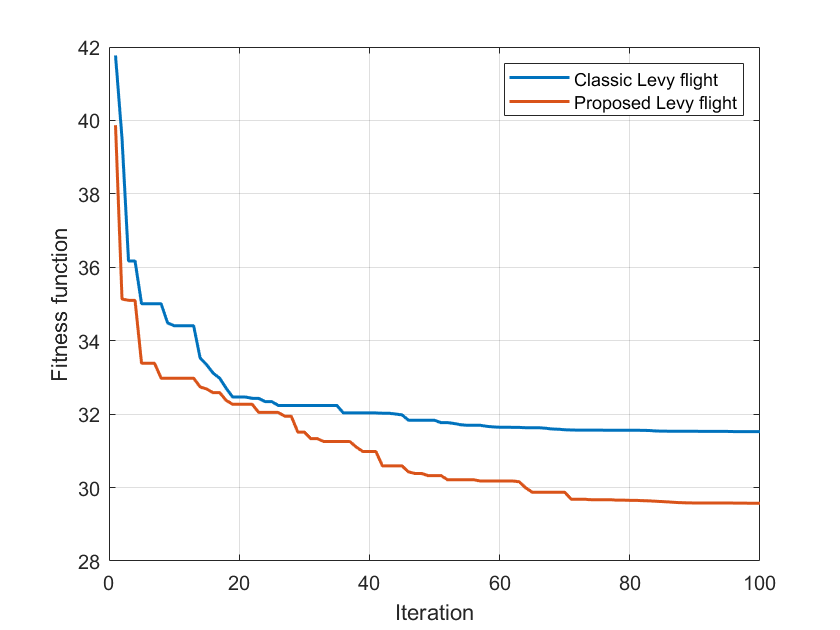}
	\caption{Action refinement using different Lévy flight searches}
	\label{fig:fig7}
\end{figure}

Moreover, Lévy flight search is applied to refine the action in the proposed framework.
Fig. \ref{fig:fig7} characterizes the action refinement performance using classic Lévy flight search and our proposed Lévy flight search. It can be observed from Fig. \ref{fig:fig7} that the proposed Lévy flight search achieves better performance with lower fitness value than classic Lévy flight search. The reason of higher accuracy of the proposed Lévy flight search is that the channel state information is applied to guide the action exploration and the Lévy flight crossover is applied to jump out from the local optimum during the search process.

\begin{figure}[htpb]
	\centering
	\includegraphics[width=8.8cm]{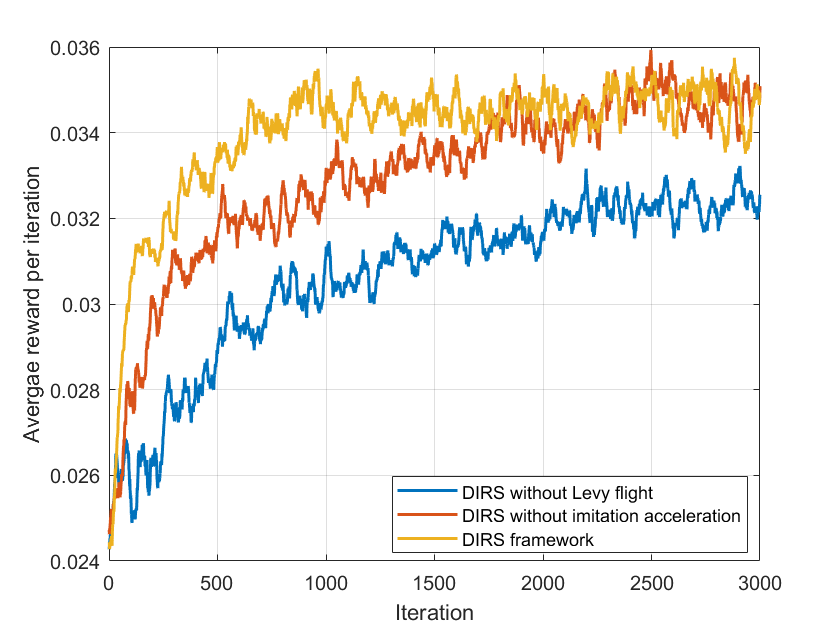}
	\caption{Average reward curve}
	\label{fig:fig9}
\end{figure}

Next, we analyze the benefits of Lévy flight search and imitation acceleration for the whole DIRS framework by using the performance metric of the average reward during the training process. One can see from Fig. \ref{fig:fig9} that the DIRS with Lévy flight search converges to a higher average reward than DIRS without Lévy flight search. This is due to the fact that the Lévy flight search is a heuristic local search which can refine the action and jump out of the local optimum by Lévy steps. One can also see that the DIRS with imitation acceleration converges faster than DIRS without imitation acceleration. This is because the imitation learning with demonstration data accelerates the convergence speed of DIRS.

Finally, we evaluate the DIRS framework with 2 well-known policy-based DRL algorithms including Actor-Critic and DDPG. TABLE \ref{tab:table2} characterizes the Training time, Inference time and Average reward of all DRL methods for online joint resource scheduling. It can be observed that the DIRS framework achieves the highest average reward while consuming the least training time. The superiority of the DIRS framework can be attributed to three aspects: (1) Imitation learning from the demonstration data accelerates the training process of DIRS; (2) Ensemble learning simplifies the structure of DNN by state space partition, which leads to less learning time of each agent, while decision consolidation according to the global information improves the reward of DIRS; (3) Lévy flight search refines the action and enhances the exploration, which leads to efficient search and fast convergence. It also can be seen the DIRS framework has shortest inference time, which can be explained for the following reasons: (1) The DIRS is a distributed DRL framework, in which the trained agents can make offloading decisions in a parallel way; (2) Each agent just needs to solve the local optimization problem in Eq. (\ref{eq:Shi17}), which is simpler than the original optimization problem.
\begin{table}[]
	\centering\makegapedcells
	\caption{The offloading performance comparison of different DRL scheduling strategies.}	
	\label{tab:table2}
	\begin{tabular}{|p{40pt}<{\centering}|p{55pt}<{\centering}|p{55pt}<{\centering}|p{50pt}<{\centering}|p{45pt}<{\centering}|}
		\hline 
		Metric &Training time (Sec) &Inference time (Sec)  & Average reward \\ 
		\hline 
		DIRS framework& 2970.84 & 0.0165  & 0.0348  \\ 
		\hline 
		Actor-Critic & 3482.72 & 0.0195  &  0.0309\\ 
		\hline 
		DDPG& 3823.15 & 0.0192  & 0.0317 \\ 
		\hline 
	\end{tabular}
\end{table}

\subsection{Decentralized inference performance evaluation}

The inference performance of the DIRS framework is compared with the following benchmark methods:
\begin{itemize}	
	\item Random offloading (Random) denotes that the offloading decision is decided randomly for each IoTD. If the computational resource of the allocated MEC is insufficient, IoTD executes the task locally.
	
	\item Greedy offloading (Greedy) denotes that all the IoTDs offload the tasks to the nearest MEC. If the computational resource is insufficient, the IoTDs who need more computing resources execute the task locally.
	
	\item Local execution (Local) denotes that all IoTDs decides to execute the task locally.		
\end{itemize}

\begin{figure}[htpb]
 	\centering
 	\includegraphics[width=8.8cm]{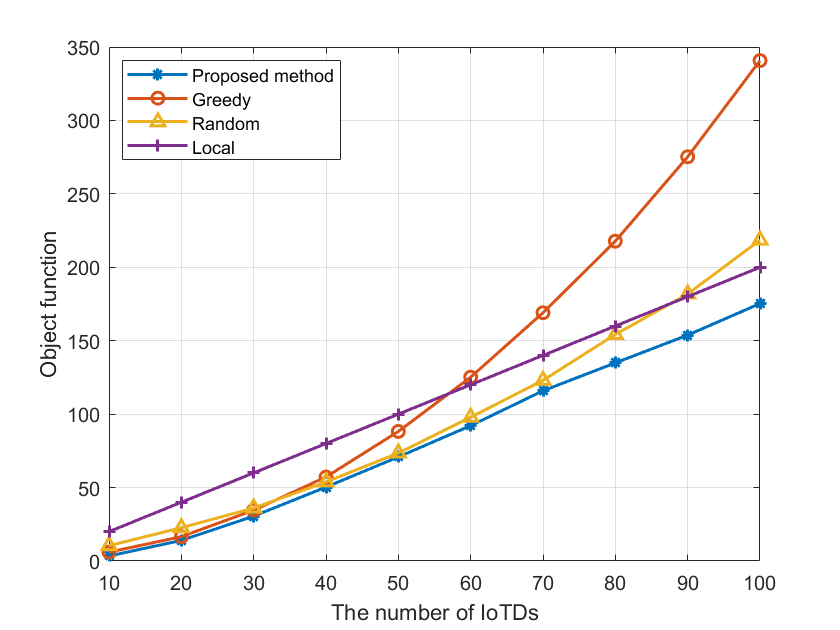}
 	\caption{The comparison of the object function when the number of IoTDs changes from 10 to 100}
 	\label{fig:fig10}
 \end{figure}

In Fig. \ref{fig:fig10}, we compare the average time and energy consumption between the proposed framework, Greedy offloading, Random offloading and Local execution, with the number of IoTDs varying from 10 to 100. As shown in Fig. \ref{fig:fig10}, the average time and energy consumption of all the methods increase gradually with the number of IoTDs increases. One can see that the proposed method achieves the best performance, especially when the number of IoTDs is large.

  \begin{figure}[htpb]
	\centering
	\includegraphics[width=8.8cm]{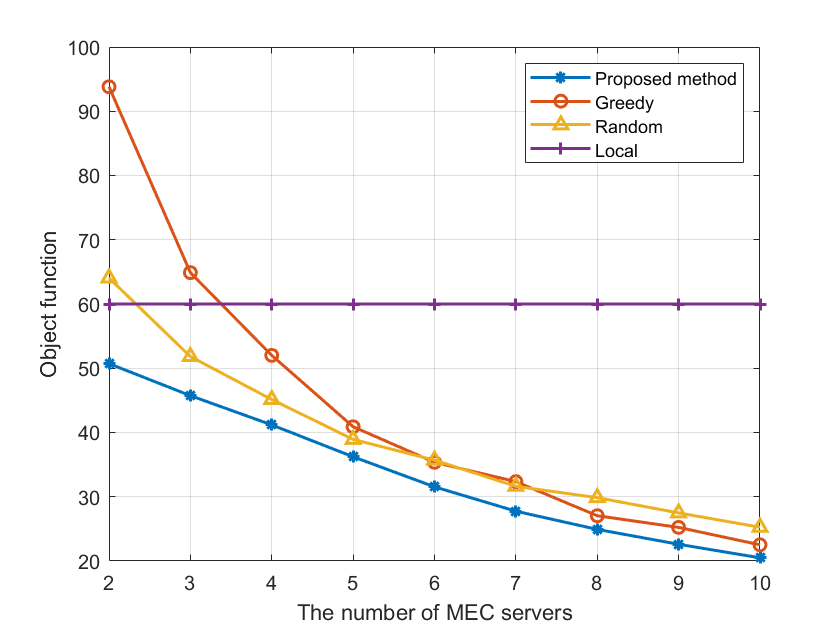}
	\caption{The comparison of the object function when the number of MEC servers changes from 2 to 10}
	\label{fig:fig11}
\end{figure}
Fig. \ref{fig:fig11} shows the average time and energy consumption achieved by the proposed method, Greedy, Random and Local, where the number of MEC servers changes from 2 to 10. One can see that the proposed method also achieves the best performance. It is demonstrated that the proposed method is capable of optimizing the offloading decision and resource allocation jointly at high accuracy, making real-time resource scheduling that are feasible for large-scale MEC systems.

\section{Conclusion}
In this paper, a novel DIRS framework has been proposed for large-scale MEC systems. This framework adopts a distributed DRL to jointly optimize computation offloading, transmission power and recourse allocation, with the objective of minimizing the sum of task latency and energy consumption for all the IoTDs. More specifically, the proposed DIRS framework consists of a multi-agent ensemble assisted DRL architecture for large state space partition and decision consolidation, a Lévy flight search for action refinement, and an imitation acceleration for agent pre-training process. The simulation results demonstrate that the DIRS framework has better performance than the existing benchmarks, and it exhibits enormous potential in the large-scale application scenarios.


%

%




\bibliographystyle{ieeetran}
\bibliography{bare_jrnl_bobo}
\newpage
\end{document}